\documentstyle[sprocl]{article}

\def\Journal#1#2#3#4{{#1} {\bf #2}, #3 (#4)}

\def\IJMP{{\em Int. J. Mod. Phys.} E}

\def\NPB{{\em Nucl. Phys.} B}
\def\PLB{{\em Phys. Lett.}  B}

\def\PRL{\em Phys. Rev. Lett.}
\def\PRE{{\em Phys. Rev.} E}
\def\PRD{{\em Phys. Rev.} D}
\def\ZPC{{\em Z. Phys.} C}

\def\PRC{{\em Phys. Rep.} C}

\def\ra{\rightarrow}

\def\be{\begin{equation}}
\def\ee{\end{equation}}
\def\bea{\begin{eqnarray}}
\def\eea{\end{eqnarray}}

\begin{document}

\title{NEW FAMILIES OF SCALING MULTIPARTICLE DISTRIBUTIONS}

\author{R. BOTET}

\address{LPS, B\^atiment 510, URA2, Universit\'{e} Paris-Sud, Centre d'Orsay, \\F-91405 Orsay,
France\\E-mail: botet@lps.u-psud.fr}

\author{M. PLOSZAJCZAK}

\address{Grand Acc\'{e}l\'{e}rateir National d'Ions Lourds, BP 5027,\\ 
F-14076 Caen, France \\E-MAIL: ploszajczak@ganac4.in2p3.fr}         



\maketitle\abstracts{Recently equations for the 
generating functional in the perturbative quantum
chromodynamics (QCD) have been extended by including the non-perturbative 
dissipation in QCD jets. The resulting equations
have been solved {\it rigorously} and new
family of scaling solutions , the so-called $\delta $ - scaling, generalizing 
the well-known Kubo-Nielsen-Olesen scaling law
for hadron multiplicity distributions
have been found. The relevance of $\delta $- scaling 
is discussed in the 
Landau - Ginzburg theory of phase transitions.
Preliminary application of these ideas to the $p{\bar p}$ data of the UA5
Collaboration is presented.}

\section{Introduction} 
A suitable framework for the comparison of multiplicity distributions at
different energies is provided by the idea of KNO scaling \cite{koba}, which
states that at sufficiently high energies :
${\lim}_{s \ra \infty} (<n>\sigma_n)(\sum_{n}^{}\sigma_n)^{-1} = f(z)$,
where $\sigma_n$~ is the cross-section for producing exactly 
$n$ particles,
$<n>$~ is the average multiplicity of produced particles and $f(z)$~ is an
energy-independent scaling function in $z = {n}/{<n>}$. 
The KNO scaling is fullfilled in
$e^{+}e^{-}$ collisions up to the highest energies considered \cite{delphi} .
The $e^{+}e^{-}$  data are well accounted for by the
Perturbative Quantum Chromodynamics (PQCD) \cite{webber} calculations and,
in particular, by the Fragmentation - Inactivation - Binary  (FIB) 
process \cite{kno,sing1}, which is equivalent to PQCD with 
dissipation effects in jet cascading\cite{zphys} .
In $p{\bar p}$ collisions, KNO scaling holds up to ISR
energies but, at higher energies, UA5 Collaboration reported 
significant breaking of the KNO scaling \cite{ua5} .

The KNO scaling was derived
from the hypothesis of Feynman scaling \cite{koba}. 
It is now known that Feynman scaling is violated at very high energies so
other arguments should be looked for. In this context,
a possible relation between the KNO scaling and the phase 
transition in Feynman-Wilson gas \cite{antoniou} 
as well as the criticality of self-similar
FIB process \cite{kno,sing1} was pointed out.
It seems therefore that KNO scaling
may have more profound reasons to appear. Since new 
data on multiplicity distributions at still higher energies will be available 
soon , we believe that the discussion of
information contained in multiplicity distributions is particularly urgent and
challenging.

\section{Order parameter fluctuations}\label{sec:order} 
Let us consider a self-similar statistical system of size $N$, 
in which phase changes are
characterized by the order parameter $\eta$. Self-similarity 
in such sytem means that for different scales which are 
characterized by different values
of order parameter $<\eta >$, fluctuations of 
normalized order parameter $\eta /<\eta >$ are {\it identical}.
We know that this is a feature of thermodynamical systems such as the
Ising model at the critical point of second order phase transition. 
Defining the anomalous dimension as : 
\begin{eqnarray}
\label{anomdim}
g =  {\lim}_{N \rightarrow \infty} \frac{d}{d\ln N} \left( 
\ln <N|\eta |> \right) ~~~\ , 
\end{eqnarray}
one can show that the partition function : 
$Z_N \sim N^{-(1-g)} \sim <|\eta |>$ and the probability density 
$P[\eta ]$~ obeys the scaling law :
\begin{eqnarray}
\label{firstsc}
<|\eta |> P[\eta ] = \Phi (z_{(1)}) =  \Phi \left( 
\frac{\eta - \eta^{*}}{<|\eta |>} \right) ~~~ \ ,
\end{eqnarray}
in the centered variable $z_{(1)} =(\eta-\eta^{*})/<|\eta|>$, with $\eta^{*}$ the most 
probable value of $\eta$ for given parameters of the system.             
We call (\ref{firstsc}) {\it the first scaling law}.
The scaling limit is defined by the asymptotic behaviour of $P[ \eta ]$
when $\eta \rightarrow \infty$, $<| \eta |> \rightarrow \infty$, but 
$z_{(1)}$~ has a finite value.  The first scaling
law holds also for fluctuations of any power ${\eta}^{\zeta}$
($\zeta > 0$) of the order parameter \cite{prlnow}. If the
order parameter is related to cluster multiplicity, like in the FIB 
process \cite{sing1,prlnow}, 
and if the most probable value of the order parameter is close enough to its 
average value,
then (\ref{firstsc})
can be written in an equivalent form to the KNO scaling \cite{koba}~.
For large $z_{(1)}$,
the scaling function behaves as : 
\begin{eqnarray}
\label{tail0}
\Phi (z_{(1)}) \sim \exp \left( -z_{(1)}^{1/(1-g)} \right) ~~~ \ ,
\end{eqnarray}
allowing for an alternative determination of anomalous dimension. 
For equilibrium systems at
the critical point of second-order phase transition, 
$g$~ must be contained in between 1/2 and 1.

We have to ask, what happens if the observable quantity is
not an order parameter but the $N$-dependent function of it like :
$m=N^{\kappa } - N \eta $, where           
$\kappa > g$. For large $N$~, $|m|$ is of order $N^{\kappa}$~. 
Writing (\ref{firstsc}) with $m$ instead of $\eta$ 
and taking into account : $P[\eta ] d\eta =P[m]dm$~, 
one finds the generalized law :
\begin{eqnarray}
\label{delta}
<|m|>^{\delta }P[m] = \Phi (z_{({\delta})}) 
\equiv \Phi \left( \frac{m-m^{*}}{<|m|>^{\delta}} 
\right) ~~\ , ~~~~~ 
\delta = \frac{g}{\kappa} < 1 
\end{eqnarray}
, which will be called in the following {\it the $\delta$ - scaling}.
The scaling function ${\Phi} (z_{(\delta )})$ 
depends only on the scaled variable :
$z_{(\delta )} = ({m-m^{*}})/({<|m|>^{\delta}})$ . 
The scaling function ${\Phi}(z_{(\delta )})$ in (\ref{delta}) has 
{\it identical form } 
as ${\Phi}(z_{(1)})$ except for the inversion of abscissa axe. 
In particular, its tail for large $z_{(\delta )}$ has the same form :
\begin{eqnarray}
\label{tail}
\Phi (z_{(\delta)}) \sim \exp \left( -z_{(\delta)}^{1/(1-g)} \right) ~~~ \ .
\end{eqnarray}
$\delta = 1/2$ is a particular case                 
when ${\Phi}(z_{({\delta})})$ is nearly Gaussian \cite{prln} and $<m> \sim N$.
This limit, called {\it the second - scaling}, has been found
 outside of the transition line in the shattering phase \cite{nowa}.
More about the $\delta$ - scaling interested reader will find in Ref. 9. 
\begin{figure}
\vspace{7cm}
\includegraphics{delphi_fig1.ps}
\label{figg1}
\caption{ The multiplicity distribution in FIB model for $\alpha = -1/3$~
with the Gaussian inactivation (\ref{inact})~ 
($\beta = 0, c = 1, \sigma = 1$) in the scaling variables
$\delta = 0.76$ are plotted for systems of different sizes : 
$N = 2^{10}$ (circles), $2^{12}$ (squares), $2^{14}$ (crosses). $10^7$ events
have been calculated. Details of the calculation can be found in Ref. 11.}
\end{figure}

The dynamical realization of $\delta$ - scaling has been found in 
FIB model with finite-scale dissipation effects. 
For an appropriate choice of the fragmentation kernel function, FIB
 is exactly equivalent to PQCD with the dissipation in jet
fragmentation. In Fig. 1 we show the cluster multiplicity distributions
calculated in FIB model for different initial sizes and plotted in the
scaling variable $z_{(0.76)}$. The calculation has been
done for homogeneous fragmentation function : $F_{j,N-j} \sim
[j(N-j)]^{\alpha}$ and the 
dissipation function which at small scales is approximated by the
Gaussian inactivation rate function : 
\begin{eqnarray}
\label{inact}
I_k = ck^{\beta}\exp 
[-[(k-1)^2/N^2]/(2{\sigma}^{2}) ] . 
\end{eqnarray}                        
The parameters of fragmentation and inactivation kernel functions are :
$\alpha = -1/3$, $\beta = 0, c = 1, \sigma = 1$. Of course, 
the inactivation becomes scale-invariant if $\sigma \rightarrow \infty $~ and 
for this choice of $\alpha $ and $\beta $, FIB process is in the
shattering phase where the second - scaling holds. On the other hand, if
$\sigma \rightarrow 0$, then $I_k \rightarrow 0$ and the fragmentation process
becomes independent of the cluster size $k$. This limiting situation happens at
the transition line where the first-scaling holds. Thus,
$\delta$ - scaling in FIB appears rather as a cross-over phenomenon between first- and
second- scaling domains than the phenomenon of gradually increasing
deviation between the order parameter in the studied process and the relevant
observable.
Below, we shall address
this question rigorously in the Landau-Ginzburg (L-G) 
theory of phase transitions.

\section{$\delta$ scaling in the Landau-Ginzburg theory}\label{sec:model} 
Let us consider the homogeneous L-G free energy density 
as the simplest example of second-order phase transition. If 
$\epsilon$ is the relative distance to the critical point, then the free-energy
density is :
$f(\eta) = \epsilon \eta^2 + b \eta^4 + \cdots $ ,
with $b$ a positive constant. Under this form, the most probable
value of $\eta$ is implicitely set to 0 in the disordered phase. 
For finite size systems,
it is often more convenient to work with the extensive 
order parameter : ${\hat m}=N \eta$.
Keeping the first two terms in $f(\eta)$, 
one may show that the probability that the system will 
be in a state ${\hat m}$ for a given $\epsilon$ is :
\begin{eqnarray}
\label{proba}
P[{\hat m}] =\frac{1}{Z_N} \exp \left( -\epsilon \frac{{\hat m}^2}{N} - 
b \frac{{\hat m}^4}{N^3} \right) ~~~ \ ,
\end{eqnarray}
where $Z_N$ is defined by the normalization of $P[{\hat m}]$. 
Without loss of generality, we consider now the case
when $\eta$ is contained between 0 and 1, i.e., ${\hat m}>0$.
We know that at the transition point ($\epsilon =0$), 
the function $P[{\hat m}]$ verifies the first-scaling \cite{kno,prlnow} 
, while below the critical point
($\epsilon<0$), $P[{\hat m}]$ verifies the second-scaling \cite{prln}.
These results are valid in the thermodynamic limit.
How the finite system moves on
between these two scalings when the control parameter $\epsilon$
moves away from 0? To answer this question, 
let us first write down the average value of ${\hat m}$ :
\begin{eqnarray}
\label{mmoy}
<{\hat m}> = N^{3/4} \frac{\int_{0}^{N^{1/2}} u^{1/2} 
\exp(-\epsilon N^{1/2} u - b u^2) du}
{\int_{0}^{N^{1/2}} u^{-1/2} \exp(-\epsilon N^{1/2} u - b u^2) du} ~~~ \ .
\end{eqnarray}
When $N \rightarrow \infty$, then the term in brakets is only 
a function of $\epsilon N^{1/2}$. This is the finite-size scaling
of this system near its critical point. It can be rewritten formally as :
\begin{eqnarray}
\label{fss}
<{\hat m}> = N^{3/4} \psi(\epsilon N^{1/2}) 
\end{eqnarray}
At the transition point, $\psi$ has a constant value 
$\psi(0)$, and $<{\hat m}>$ behaves as a power of $N$ with the mean-field 
exponent 3/4. At the critical point, this exponent 
is identical to the anomalous dimension defined in (\ref{anomdim}). 
In the following discussion, it will be denoted by $g$ both for systems 
at the critical state and outside of it. 
In the whole ordered phase, we have $g=1$. Eq. (\ref{mmoy}) shows also 
that the cross-over between critical and non-critical phases must be discussed 
according to the values of $\epsilon N^{1/2}$. Moreover, since we 
have to recover the non-critical case : $<{\hat m}> \propto N$,
for finite negative $\epsilon$ therefore : $\psi(x) \sim (-x)^{1/2}$, when $x$ 
is large and negative.     
Let us note first the difference between the average value of 
${\hat m}$, as given
by (\ref{mmoy}), and its most probable value (for $\epsilon \leq 0$) :
${\hat m}^{*} = \pm N^{3/4} [ (-\epsilon N^{1/2})/2b ]^{1/2}$.
For $\epsilon <0$, one can approximate $P[{\hat m}]$ by a Gaussian function :
\begin{eqnarray}
\label{gaus}
\frac{N^{3/2}}{<{\hat m}>} P[{\hat m}] \sim \sqrt{\frac{4b}{\pi}} \exp \left(
{\frac{\epsilon}{N}}({\hat m}-{\hat m}^{*})^2 \right) ~~~ \ .
\end{eqnarray} 
Suppose now that the order of magnitude of $\epsilon$ is such that
$\epsilon \simeq 1/N^{a}$, with some positive exponent $a$, smaller than
1/2 to ensure that $\epsilon N^{1/2}$ is a large number for large $N$ .
Then, $<{\hat m}>$ and ${\hat m}^{*}$ are both of similar order of magnitude. 
This means that the prefactor (${N^{3/2}}/<m>$) in
(\ref{gaus}) is of order : $<{\hat m}>^{(1+a)/(2-a)}$. In a similar way, the
quantity (${N/\epsilon}$) appearing in (\ref{gaus}) is of order :
$<{\hat m}>^{2(1+a)/(2-a)}$. 
Therefore, one can now write eq. (\ref{gaus}) in 
the $\delta$ - scaling form (\ref{delta}). Obviously,
we recover the known particular cases : $\delta=1$ for $a=1/2$, and $\delta=1/2$
for $a=0$. Here, however, the $\delta$-scaling appears as a finite-size effect
and the tail of scaling function for large
arguments has {\it always} Gaussian form, in contrast
to the $\delta$ - scaling due to $N$ - dependent change of order parameter
which was discussed  in the preceding section.
\begin{figure}
\vspace{7cm}
\includegraphics{delphi_fig2.ps}
\label{fig2}
\caption{ The UA5 multiplicity distributions in $p{\bar p}$ reactions
\protect\cite{ua5} 
for three different
energies : 200 GeV (circles), 546 GeV (squares) and 900 GeV (diamonds), are
plotted in the $\delta$ - scaling variables  for $\delta = 0.9$.}
\end{figure}

We have seen above that $\delta$-scaling appears either as a result of
incompatibility between the observable and the order parameter
, or as a cross-over between first- and second-scaling. Both $N$-dependent
phenomena may actually be related one to another.
In general, it is not so easy to conclude which of the two
cases prevails without an additional information, and this point
needs some comments here. Up to now, we did not need to know explicitely
the total size $N$ of the system, because the  
$\delta$-scaling is not expressed explicitely in terms of $N$. In
the practical applications, we have to know only that the data 
corresponds to a {\it constant value} of $N$.
Nevertheless, we could have in principle access to the $N$ - dependence of 
average quantities  such as $<m>$. It should be :
\begin{eqnarray}
\label{mfirst}
<m> \sim N^{g/ \delta} ~~~ \ ,
\end{eqnarray}
for the change of variable as in Sect. \ref{sec:order} and :
\begin{eqnarray}
\label{msecond} 
<m> \sim N^{2g/(\delta+1)} ~~~ \ ,
\end{eqnarray}
for the finite-size cross-over effect 
described above in the mean-field approximation. In the latter case,  
$g=3/4$. If one knows from the experimental data
the value of scaling parameter $\delta$ 
and the $N$-dependence of $<m>$ , then
one may find the value of the anomalous dimension if the cause of
$\delta$ - scaling can be decided by analyzing 
the tail of the scaling distribution. Below, we shall illustrate 
this on the example of UA5 data.

\section{Application and conclusions}\label{sec:concl} 
We shall apply now the concepts developed above to the $p{\bar p}$ data of UA5
Collaboration\cite{ua5} which show significant deviations with respect to the KNO
(first-) scaling. If one plots data for different ${\sqrt s}$ in KNO 
variables, one notice that the maxima of $<m>P[m]$ function are shifted , the
height of maximum increases with ${\sqrt s}$ and the width decreases.
Distributions at ${\sqrt s}=200$, 546 and 900 GeV are plotted in 
Fig. 2 in the $\delta$ - scaling variables for $\delta=0.9$. 
\begin{figure}
\vspace{7cm}
\includegraphics{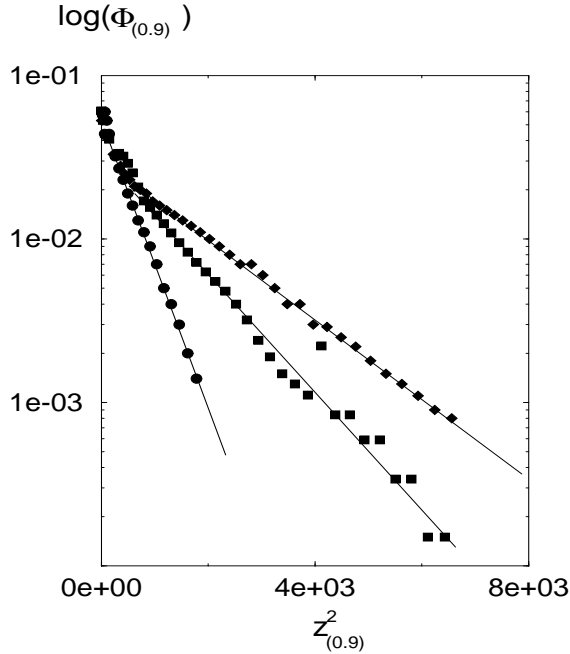}
\label{fig3}
\caption{ The logarithm of scaled multiplicity distributions at UA5 data 
\protect\cite{ua5} and three different
energies : 200 GeV (circles), 546 GeV (squares) and 900 GeV (diamonds), plotted
versus $z_{(0.9)}^2 \equiv ((m-m^{*})/<m>^{0.9})^2$ for $z_{(0.9)} >0$.}
\end{figure}
Even if uncertainty on
the average multiplicity is probably large, the scaling looks well. As
noticed before, we can find argument for the cause of $\delta$ - scaling by
analyzing the tail of scaling function. This is shown in Fig. 3
where the logarithm of scaling distribution is plotted versus 
$z_{(0.9)}^{2}$. We see that the three curves tend to be linear in
this plot, proving that tails of ${\Phi}(z_{(\delta )})$ 
are essentially Gaussian. This provides a rather convincing argument that 
the fragmenting system is predominantly 
in the ordered phase ($\delta<1$, and Gaussian tail),
close to a critical point ($\delta \simeq 1$). Moreover, a plot 
(unfortunately, for  only three values of $\sqrt{s}$) 
of $<m>$ vs $\sqrt{s}$ shows that : $<m> \sim (\sqrt{s})^{0.35}$. 
Using relation (\ref{msecond}), allows then to write : 
\begin{eqnarray}
\label{finisz}
N^g \sim (\sqrt{s})^{0.33} ~~~ \ , 
\end{eqnarray}
what should be the proper scaling of the order parameter with the size of 
the system. Of course, one should be aware of the preliminary character of this
extracted value of $g$. First of all, data in full space are not numerous.
Secondly, the analysis is global in the sense that it concerns 
different classes of events mixed together. Nevertheless, the $\delta$ -
scaling analysis discussed in this work 
may become the powerful tool, allowing for an intelligent analysis of
multiplicity data at different energies, which could give access to the
determination of the anomalous dimension 
in the particle production at ultrarelativistic energies.

\vskip 1truecm
{\bf Acknowledgements} \\
We would like to thank B. Buschbeck, S. Hegyi and A. Giovannini for interesting discussions.


\end{document}